\def\autoeq{ {\global\advance\count90 by 1} \eqno(\the\count90) }
\def\autoref{ {\global\advance\refno by 1} \kern -5pt [\the\refno]\kern 2pt}
\def\setup{\count90=0 \count80=0 \count91=0 \count85=0
\countdef\refno=80 \countdef\secno=85 \countdef\equno=90
\countdef\ceistno=91 }

\def\Box{\vbox{\hrule
 \hbox{\vrule height 6pt \kern 6pt \vrule height 6pt}\hrule}\kern 2pt}
\setup
\overfullrule=0pt
\magnification1200
\noindent
\hsize 130mm
\centerline{  }
\rightline{DIAS-STP 97-23}
\rightline{21st October 1997}
\vskip 2cm
\centerline{\bf Renormalisation Flow and Geodesics on the Moduli Space}
\centerline{\bf of Four Dimensional $N=2$ Supersymmetric Yang-Mills Theory} 
\vskip 1.2cm
\centerline{Brian P. Dolan}
\vskip .5cm
\centerline{\it Department of Mathematical Physics}
\centerline{\it National University of Ireland, Maynooth, Ireland}
\centerline{and}
\centerline{\it Dublin Institute for Advanced 
Studies}
\centerline{\it 10, Burlington Rd., Dublin, Ireland}
\vskip .5cm
\centerline{e-mail: bdolan@thphys.may.ie}
\vskip 1.5cm
\baselineskip=1.5 \baselineskip
\centerline{\bf ABSTRACT}
\bigskip
\noindent It is shown that the $\beta$-functions for  
four dimensional $N=2$ supersymmetric Yang-Mills theory without matter
give integral curves on the moduli space some of which are geodesics of the 
natural metric on the moduli space. 
In particular the flow lines which cross-over from
from the weak coupling limit (asymptotically free theory) to the singular
points, representing the strong coupling limit, are geodesics. A possible
connection with irreversibility is discussed. 
\smallskip

\vfill\eject

The purpose of this letter is to investigate the geometrical and 
topological structure of the renormalisation flow in an exactly solved four dimensional field 
theory - $N=2$ supersymmetric $SU(2)$ Yang-Mills theory without matter,
\autoref\newcount\SW\SW=\refno.  
The $\beta$-function for the complex coupling, 
$\tau={\theta_{eff}\over 2\pi}+{4\pi i\over g_{eff}^2}$, of this model was 
calculated in 
\autoref\newcount\MiNem\MiNem=\refno
\autoref\newcount\Matone\Matone=\refno
 and the resulting flow diagram is shown
in figure 1.
This diagram has special points: the weak coupling limit
$\tau=i\infty$, the strong coupling limit at $\tau=0,\pm 1$ 
where there are singularities due to extra massless degrees of freedom associated with composite objects (monopoles and dyons),
and the  points at $\tau=(\pm 1+i)/2$ where
the Higgs field $\phi$ satisfies  
$Tr\langle\phi^2\rangle = 0$.
The points at $\tau=\pm1$ and $\tau=0$ are physically
equivalent under the symmetry $\tau\rightarrow\tau +1$, as are the two 
points $\tau=(\pm 1+i)/2$.
At these last two points $\langle\phi\rangle\ne0$ even though 
$Tr\langle\phi^2\rangle = 0$, so that $SU(2)$ symmetry is not restored in the quantum regime, but these points are the quantum mechanical vestige
of the place where the symmetry would be restored in the classical theory.  
It will be shown that the flow lines which cross over between all these 
special points are geodesics of the Seiberg-Witten metric, but none of the other flow lines is a geodesic.  These are geodesics with a \lq\lq frictional" force and the geodesic nature of the cross-over appears to be related to irreversibility of the renormalisation flow. 
\bigskip
$N=2$ supersymmetric Yang-Mills theory has features in common with both the standard model of electro-weak theory (symmetry breaking and mass generation by the Higgs mechanism) and with quantum chromo-dynamics (asymptotic freedom and confinement), [\the\SW]. 
There are also many differences, of course, but as a model for testing physical ideas in four dimensions Seiberg and Witten's low energy effective action gives us an unprecedented arena for experimentation.  In common with $QCD$,   $N=2$ supersymmetric Yang-Mills theory exhibits dimensional transmutation in that there exists a natural length scale, $\Lambda$, at which the interactions become strong in the sense that the dimensionless 
gauge coupling, $g_{eff}(\Lambda)$, becomes large.  It is convenient,
as above, to combine $g_{eff}(\Lambda)$ with an effective 
topological parameter, 
$\theta_{eff}(\Lambda)$, into the complex quantity $\tau = {{\theta_{eff}}\over{2\pi}} + {{4\pi i}\over{g_{eff}^2}}$.  The value of $\tau$ depends on Higgs expectation  value, $\langle\phi\rangle=\left(\matrix{a&0\cr 0&-a\cr}\right)$, which is not fixed in this theory but can take on a range of values corresponding to different vacua. Using the gauge invariant quantity 
$u = Tr\langle\phi^2\rangle$ to parameterise the symmetry breaking, Seiberg and Witten determined the zero momentum functional form of the dimensionless quantity $\tau$ as a function of $u/ \Lambda^2 \equiv \tilde u$,
$$
\tau = {{i K^\prime}\over{K}}
\autoeq
$$
\newcount\taudef\taudef=\equno
\medskip
\noindent 
where $K^\prime(k^2) = K(k^{\prime 2})$ and $K(k^2)$ are standard elliptic integrals 
\autoref,\newcount\WW\WW=\refno
with $k^{\prime 2} = 1 - k^2$ and $k^2 = {{2}\over{\tilde u + 1}}$
(the explicit form (\the\taudef) can be found in 
the review
\autoref\newcount\AGH\AGH=\refno).  
It is then
straightforward to obtain the $\beta$-functions of the theory, including non-perturbative effects, [\the\MiNem] [\the\Matone],
\medskip
$$
\beta(\tau):= \quad \Lambda{{\partial\tau}\over{\partial\Lambda}}\bigg\vert_u 
=\quad -2u{{\partial\tau}\over{\partial u}}\bigg\vert_\Lambda 
=\quad {1\over i\pi}\left({{\vartheta_3^4+\vartheta_4^4}\over{\vartheta_3^4\vartheta_4^4}}\right).\autoeq
$$ 
\newcount\betafun\betafun=\equno
\noindent
The $\vartheta$-functions used here are, in the conventions of Whittaker and Watson [\the\WW],
$$\vartheta_2 = 2q^{1/4} \prod_{n=1}^\infty (1-q^{2n})\,(1+q^{2n})^2$$
$$\vartheta_3 = \prod_{n=1}^\infty (1-q^{2n})\,(1+q^{2n-1})^2$$
$$\vartheta_4 = \prod_{n=1}^\infty (1-q^{2n})\,(1-q^{2n-1})^2
\autoeq
$$
\newcount\thefun\thefun=\equno
with $q = e^{\pi i\tau}$.
\bigskip
These $\beta$-functions represent a vector flow,
$\vec\beta=\beta{\partial\over\partial\tau}+
\bar\beta{\partial\over\partial\bar\tau}$,
 on a one (complex) dimensional manifold, parameterised by $\tau$, which has the topology of a sphere with three holes 
[\the\SW].  The flow is most easily visualised by noting that it is radial in the $\tilde u$-plane (by definition) and then transforming to the $\tau$-plane.  The $\beta$-function is singular at $\tau=\pm 1,0$ and vanishes at $\tau={1\over 2}(\pm 1+i)$.  This last point corresponds to $u=0$, which is the point at which one would have $a=0$ classically, restoring the gauge symmetry to the full $SU(2)$, but quantum mechanically $\vert a \vert$ has a lower bound ($\approx  0.7628$) so $a = 0$ can never be reached in the quantum regime.  Nevertheless the point $u=0$ corresponds to the minimum value of $\vert a \vert$ in the quantum theory
\autoref\newcount\Ferrari\Ferrari=\refno.
The $\beta$-function (\the\betafun) diverges at the points $\tau=0,\pm1$,
but the invariant length squared $|\beta |^2=G_{\tau\bar\tau}\beta\bar\beta$
vanishes at these points, thus they can be thought of as fixed points
within the approximations used by Seiberg and Witten.
\bigskip
Note that this vector flow represents a real physical flow that follows from changing $\Lambda$ at fixed $u$ $,(\hbox{or}\,\, u$ at fixed $\Lambda$).  
It is thus similar in spirit to  
the Callan-Symanzik flow of Q.E.D., which one obtains by changing the physical electron mass.  Here physical masses are also being varied, since $u$ (and thus $a$) is changing at fixed $\Lambda$.
\bigskip
The question posed here is, how does the above flow relate to the geometry of the punctured sphere?  Seiberg and Witten also described a metric on the punctured sphere which is compatible with the $\Gamma(2)$ action on the upper 
half $\tau$-plane.  In terms of the Higgs expectation value, $a$,
$$
ds_{S-W}^{2} = (Im \,\tau) \,da d\bar a.
\autoeq$$
Using $\tau$ co-ordinates this is
$$
ds_{S-W}^{2} = \pi^2(Im\, \tau) \bigg\vert {{\vartheta_3^4\vartheta_4^4}\over{\vartheta_2^2}}\bigg\vert^2 d\tau d\bar\tau,
\autoeq$$
\newcount\SWmetric\SWmetric=\equno
or in $u$-co-ordinates
$$
ds_{S-W}^{2} = {{1}\over{\pi^2}} {{(K^\prime\bar K+\bar K^\prime K)}\over{\sqrt{1+u} \sqrt{1+\bar u}}} du d\bar u.
\autoeq$$
The relationship between the Seiberg-Witten metric above and the Poincar\'e metric on the upper half-plane,
$$
ds_P^2 = {{1}\over{(Im \,\tau)^2}} d\tau d \bar\tau,
\autoeq$$
was discussed in 
\autoref.\newcount\MatonE\MatonE=\refno
\bigskip
Physically the Seiberg-Witten metric seems to be the more relevant - it gives rise to a positive Ricci curvature which diverges at $\tau = \pm 1$ and $0$, reflecting the singularities at these points,
$$
R = {{1}\over{\pi^2(Im\tau)^3}} 
\bigg\vert{{\vartheta_2^2}\over{\vartheta_3^4\vartheta_4^4}}\bigg\vert^2\quad,
\autoeq$$
and $R$ tends to zero in the weak coupling limit, $\tau \rightarrow i\infty$.  The boundaries of the fundamental region (figure 1) are clearly geodesics of the Poincar\'e metric and it is natural to ask if this property also 
holds for the Seiberg-Witten metric.  
A detailed calculation shows that it does!  Indeed the only flow lines of the figure that are geodesics of the Seiberg-Witten metric are the bold lines which cross-over between the 
special points $\tau=\pm 1$, $\tau = 0$, $\tau = i\infty$ and $\tau = {1\over 2}\bigl(\pm 1 + i\bigr)$.
\bigskip
To prove that the lines $\tau = is$, $\tau = 1+is$, $\tau = {1\over 2}+is$ 
(where $s$ is a real parameter) are geodesics, consider the geodesic equation for a vector field $\vec\xi = \xi^\tau{{\partial}\over{\partial\tau}} + \xi^{\bar\tau}{{\partial}\over{\partial\bar\tau}}$,
$$
\nabla_{\vec\xi} \vec\xi = \lambda \vec\xi.
\autoeq$$
Explicitly, for the Levi-Civita connection associated with a K\"ahler metric,
$$\eqalign
{\xi^\tau{{\partial\xi^\tau}\over{\partial\tau}} +
\xi^{\bar\tau}{{{\partial\xi^{\tau}}\over{\partial\bar\tau}}} &+ \Gamma_{\tau\tau}^{\tau}\xi^\tau\xi^\tau = \lambda\xi^\tau\cr
\xi^\tau{{{\partial\xi^{\bar\tau}}\over{\partial\tau}}} +
\xi^{\bar\tau}{{{\partial\xi^{\bar\tau}}\over{\partial\bar\tau}}} &+ \Gamma_{\bar\tau\bar\tau}^{\bar\tau}\xi^{\bar\tau}\xi^{\bar\tau} = \lambda\xi^{\bar\tau}.\cr}
\autoeq
$$
\noindent For a holomorphic vector field, $\xi^\tau=\xi(\tau)$ and 
$\xi^{\bar\tau}=\overline{\xi(\tau)}$, with $\bar\xi$ the complex conjugate of $\xi$,
these equations reduce to
$$\eqalign
{\xi{{\partial\xi}\over{\partial\tau}} &+ \Gamma_{\tau\tau}^{\tau}\xi^2 = \lambda\xi\cr
\bar\xi{{\partial\bar\xi}\over{\partial\bar\tau}} &+ \Gamma_{\bar\tau\bar\tau}^{\bar\tau}\bar\xi^2 = \lambda\bar\xi,\cr}
\autoeq
$$
\newcount\geodesic\geodesic=\equno
\noindent with the condition that $\lambda = {{\partial\xi}\over{\partial\tau}} + \Gamma_{\tau\tau}^{\tau} \xi = {{\partial\bar\xi}\over{\partial\bar\tau}} + \Gamma_{\bar\tau\bar\tau}^{\bar\tau} \bar\xi$ be real.  For the $\beta$-function above
and the connection for the Seiberg-Witten metric this is easily checked
using the representation (\the\thefun) for the $\vartheta$-functions.
Since $q = e^{-\pi s}$ along $\tau = is$,  $q =-e^{-\pi s}$ 
along $\tau=1+is$ and $q=ie^{-\pi s/2}$ along $\tau= {1\over 2} + is$, 
the reality conditions on $\lambda$ are readily verified using (\the\betafun)
and (\the\SWmetric).  For the circular arch $\tau = {1\over 2} (1 + e^{i\phi})$ corresponding to $-1<u<1$, a proof can be constructed using the $u$-coordinate, in which one only need prove that 
\medskip
$$
\Gamma_{uu}^u = {{\{\bar K^\prime(\partial_{u}K) + 
\bar K(\partial_{u}K^\prime)\}}\over{K^\prime\bar K + \bar K^\prime K}} - 
{{1}\over{2(u+1)}}
\autoeq
$$
is real along the real line segment $-1<u<1$.  This follows by noting that $k^{\prime 2}={{u-1}\over{u+1}}$is real and negative for $-1<u<1$, thus $K^\prime$ is real on this segment, $K$ however is not real, but one can show using (see ref. [\the\WW])
$$
{{dK}\over{d(k^2)}} = {{E-k^{\prime 2}K}\over{2k^2k^{\prime 2}}} \quad , \quad {{dK^\prime}\over{d(k^2)}} = {{k^2K^\prime - E^\prime}\over{2k^2k^{\prime 2}}},
\autoeq
$$
$$
K^\prime E+E^\prime K - K^\prime K=\pi/2,
$$
\medskip\noindent
and the fact that $K^\prime$ is real, that $\Gamma_{uu}^u$ is in fact real along $-1<u<1$, and thus this line segment is a geodesic.  

One can construct an alternative proof that the vertical
boundary lines of the fundamental domain in figure 1
are geodesics, just using group theoretical arguments rather than
heavy handed analysis. Note that the discrete map $\tau\rightarrow-\bar\tau$
is an isometry of the Seiberg-Witten metric (\the\SWmetric), since under
this map $Im(\tau)\rightarrow Im(\tau)$ and $q\rightarrow\bar q$.
This not an element of $\Gamma(2)$, the subroup of $Sl(2,{\bf Z})$
acting on the upper-half $\tau$ plane consisting of matrices which
are the identity matrix mod 2, but it is still an isometry. 
Thus the group of isometries is actually
larger than $\Gamma(2)$. Under this map, the positive imaginary axis
is in invariant and is thus a geodesic of the Seiberg-Witten metric,
since any line of fixed points of an isometry is a geodesic.
This further implies that that the vertical lines above $\tau=\pm 1$
are also geodesics, since $T:\tau\rightarrow\tau+1$ is also an isometry
(this simply changes the sign of $q$ and so interchanges
$\theta_3$ and $\theta_4$ in (\the\thefun) and multiplies $\theta_2$
by a phase, thus leaving the Seiberg-Witten line element (\the\SWmetric)
unchanged). 

The conclusion is that the bold lines in figure 1 are geodesics 
in the Seiberg-Witten 
metric. That none of the other flows is geodesic follows from the 
fact that they are all repulsed from the singular points $\tau= \pm1,0$.  
Since these points have infinite positive curvature any geodesic 
would necessarily be attracted to and focused into these points, 
not repulsed by them.
\bigskip
One can actually prove a slightly stronger result.  The two form 
$$d\beta = \partial_{[a}(G_{b]c}\beta^c)dx^a\wedge dx^b\autoeq$$
\noindent does not vanish in general  (the flow is not gradient), 
but it does vanish along the thick lines of the figure. On a K\"ahler
manifold this is equivalent to saying that 
$\partial_\tau(G_{\tau\bar\tau}\beta^\tau)$ is real on these lines. 
Combined with the observation that the $\beta$-functions are 
conformally Killing for the Seiberg-Witten metric it can be shown that
this in fact implies that these lines are geodesics.
\bigskip
It seems possible that these observations may be related to irreversibility of the flow.  Gradient flow, with a positive definite metric\autoref,
\newcount\WZ\WZ\refno
automatically implies a  c-theorem 
\autoref.
\newcount\Zamolodchikov\Zamolodchikov=\refno
If the flow is gradient then there exists a differentiable function, $U(g)$,
such that $\beta^a=G^{ab}\partial_bU(g)$ and so 
$${dU(g)\over d\ln\Lambda}=\beta^a\partial_aU(g)=\beta^a\beta^bG_{ab}\ge 0
\autoeq$$ for a positive definite metric. The function $U(g)$ is therefore
monotonic along the flow trajectories.
The flow here is not gradient with the Seiberg-Witten metric in general, but the two form $d\beta$ does vanish along the lines of cross-over, which then implies that these lines are geodesics.  Geodesic cross-over, when the metric is obtained as the second derivative of the effective potential, has been noted in some other models (specifically Gaussian models in $2\le d\le 4$ and the  $1-d$ Ising model
\autoref\newcount\GD\GD=\refno, and the $O(N)$ model in $3-d$ for large $N$
\autoref\newcount\ON\ON=\refno)
where it can be related to a maximisation of the relative entropy.  
Thus it seems plausible that geodesic flow is related to an increase in relative entropy along the lines of crossover.  
Note that geodesic flow does not imply reversibility - the geodesic flow 
described here is more closely analogous to motion under friction. 
The \lq\lq frictional''
force along the trajectories is represented by the function $\lambda$
in equation (\the\geodesic). In the weak coupling regime
$\lambda\rightarrow -1$ while at the singular points
$\lambda\rightarrow -\infty$, indicating that the frictional
force becomes infinite at these points.
It has been checked
numerically that $\lambda$ is always negative along the thick
lines of figure 1 - in line with the concept of irreversibility
in the infra-red (small $u$) direction.

It would obviously be of interest to explore these matters 
in the case of supersymmetric theories with matter and for more
general gauge groups, but the absence of such concrete formulae as equation 
(\the\taudef), would probably render the pedestrian analysis presented
 here inadequate to the task - one would need more sophisticated mathematical techniques.

It is a pleasure to thank the Alexander von Humboldt foundation for 
financial support and the Physikalisches Institut, University of Bonn,
where the bulk of this work was completed,
for their hospitality.
The author also acknowledges partial sponsorship from
Baker Consultants Ltd., Ireland, networking specialists
(http://www.baker.ie).
\vfill\eject
{\bf References}
\medskip
\item{[\the\SW]} N. Seiberg and E. Witten, Nuc. Phys. {\bf B426} (1994) 19 (hep-th/9407087)
\smallskip
\item{[\the\MiNem]} J.A. Minahan and D. Nemeschansky, Nucl. Phys. {\bf B468} (1996) 72 (hep-th/9601059)
\smallskip
\item{[\the\Matone]} G. Bonelli and M. Matone, Phys. Rev. Lett. {\bf 76} (1996) 4107 (hep-th/9602174)
\smallskip
\item{[\the\WW]} E.T. Whittaker and G.N. Watson, {\sl A Course of Modern Analysis}, C.U.P. (1940)
\item{[\the\AGH]} L. Alvarez-Gaum\'e and S.F. Hassan, Fortsch.Phys. {\bf 45} (1997) 159 (hep-th/9701069)
\smallskip
\item{[\the\Ferrari]} F. Ferrari and A. Bilal, Nucl. Phys. {\bf B469} (1996) 387 (hep-th/9602082)
\smallskip
\item{[\the\MatonE]} M. Matone, Phys. Rev. {\bf D53}, (1996) 7354 (hep-th/9506181)
\item{[\the\WZ]} D.J. Wallace and R.K.P. Zia, Phys. Lett. {\bf 48A} (1974) 325; \hfill\break Ann. Phys. {\bf 92} (1975) 142
\smallskip
\item{[\the\Zamolodchikov]} A.B Zamolodchikov, Pis'ma Zh. Eksp. Teor. Fiz. 
{\bf 43} (1986) 565
\smallskip
\item{[\the\GD]} B.P. Dolan, Int. J. Mod. Phys. {\bf A12} (1997) 2413 
(hep-th/9511175)
\smallskip
\item{[\the\ON]} B.P. Dolan, {\sl Geodesic Renormalisation Group Flow in the
$O(N)$ Model for Large $N$}, University of Edinburgh preprint 97/1 
(hep-th/9702156)
\smallskip
\vfill\eject
\includegraphics{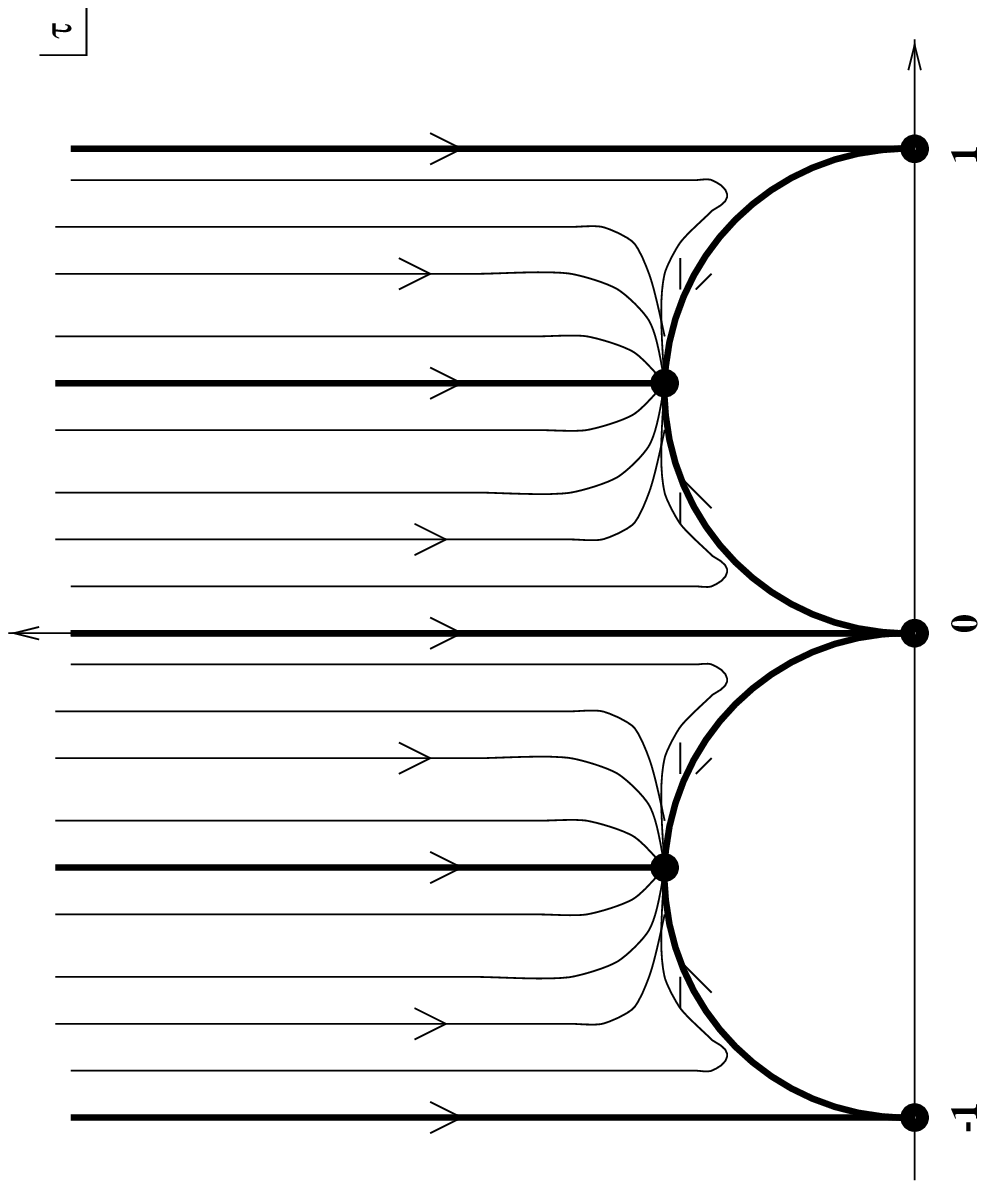}
\centerline{Fig. 1: Renormalisation flow for $SU(2)$ $N=2$ supersymmetric Yang-Mills}
\centerline{The thick lines bounding the fundamental domain are geodesics}

\end